\documentclass[12pt]{iopart}

\usepackage{graphicx}
\usepackage {amssymb}
\begin{document}

\title[Crossing-point effect in heat-capacity curves of ruthenocuprates]
{Characteristic crossing point ($T_{\ast}\approx 2.7$~K) in
specific-heat curves of samples
RuSr$_2$Gd$_{1.5}$Ce$_{0.5}$Cu$_2$O$_{10-\delta}$ taken for
different values of magnetic field}

\author{B. I. Belevtsev$^{1}$\footnote[1]{To
whom correspondence should be addressed
(belevtsev@ilt.kharkov.ua)},
V. B. Krasovitsky$^{1}$,  D. G. Naugle$^{2}$, \\
K. D. D. Rathnayaka$^{2}$, G. Agnolet$^{2}$, I. Felner$^{3}$}

\address{$^{1}$ B. Verkin Institute for Low Temperature
Physics \& Engineering, National Academy of Sciences, Kharkov,
61103, Ukraine}
\address{$^{2}$ Department of Physics, Texas A\&M University, College Station, TX 77843,
USA}
\address{$^{3}$ Racah Institute of Physics, The Hebrew University,
Jerusalem, 91904, Israel}

\begin{abstract}
Magnetic properties of polycrystalline samples of\\
RuSr$_2$(Gd$_{1.5}$Ce$_{0.5}$)Cu$_{2}$O$_{10-\delta}$, as-prepared
(by solid-state reaction) and annealed (12 hours at
845$^{\circ}$C) in pure oxygen at different pressure (30, 62 and
78 atm) are presented. Specific heat and magnetization were
investigated in the temperature range 1.8--300 K with a magnetic
field up to 8 T. Specific heat, $C(T)$, shows a jump at the
superconducting transition (with onset at $T\approx 37.5$~K).
Below 20 K, a Schottky-type anomaly becomes apparent in $C(T)$.
This low-temperature anomaly can be attributed to splitting of the
ground term ${^8}S_{7/2}$ of paramagnetic Gd$^{3+}$ ions by
internal and external magnetic fields. It is found that curves
$C(T)$ taken for different values of magnetic field have the same
crossing point (at $T_{\ast}\approx 2.7$~K) for all samples
studied.  At the same time, $C(H)$ curves taken for different
temperatures have a crossing point at a characteristic field
$H_{\ast}\approx 3.7$~T. These effects can be considered as
manifestation of the crossing-point phenomenon which is supposed
to be inherent for strongly correlated electron systems.
\end{abstract}
 \submitto{\JPCM}
\maketitle

\section{Introduction}
\label{int} So called, crossing-point phenomenon is one of the
interesting and still puzzling effects in strongly correlated
electron systems (see Refs. \cite{voll,mishra,macedo,eck} and
references therein). A typical example of this effect is
temperature behavior of specific-heat curves $C(T,X)$ taken at
different values of a thermodynamic variable $X$ (such as magnetic
field $H$ or pressure $P$): the curves cross at one temperature
$T_{\ast}$. This type of effect was found also not only for
thermodynamic but for dynamic quantities as well (for example, for
frequency dependent optical conductivity). More generally this
rather long ago known effect is termed {\it isosbestic point}
\cite{eck}.
\par
Known experiments revealed {\it isosbestic points} in different
systems of strongly correlated fermions like liquid He$^3$,
heavy-fermion compounds and others \cite{voll,mishra,macedo,eck}.
In particular, crossing point in $C(T,H)$ curves was found in
heavy-fermion compound CeCu$_{5.5}$Au$_{0.5}$ \cite{schlag},
semimetal Eu$_{0.5}$Sr$_{0.5}$As$_3$ \cite{fisch}, superconducting
cuprate GdBa$_2$Cu$_4$O$_8$ \cite{ho}  and manganite NdMnO$_3$
\cite{cheng}. Nevertheless,  general reasons and conditions for
realization of {\it isosbestic points} are still not so clear. The
known theoretical considerations \cite{voll,mishra,macedo,eck} are
based on rather different approaches. Available relevant
experimental data can be considered as meagre, therefore further
experimental findings of this phenomenon in different systems
should be helpful for understanding of its nature.
\par
In this study, the crossing-point effect is revealed in $C(T,H)$
curves  of polycrystalline perovskite-like
RuSr$_2$(Gd$_{1.5}$Ce$_{0.5}$)Cu$_{2}$O$_{10-\delta}$ (Ru1222-Gd).
This compound is from the known family  of ruthenocuprates
RuSr$_2$R$_{2-x}$Ce$_{x}$Cu$_2$O$_{10-\delta}$ (where R=Gd, Eu)
\cite{felner1,lorenz,awana,klamut}.  This family within range
$0.4\leq x \leq 0.8$ shows superconductivity with $T_c$ up to
$\approx 50$~K for $x=0.5-0.6$. Below $T_{WF}=$~80--100~K,
indications of weak-ferromagnetic order are found. It is believed
on these grounds that these compounds are magnetic
superconductors. Superconductivity is associated with CuO$_2$
planes, while magnetic order is thought to be connected with the
RuO$_2$ planes (see more in reviews
\cite{felner1,lorenz,awana,klamut}).
\par
In the following we shall present and discuss the crossing-point
phenomenon in Ru1222-Gd found in this study together with
indispensable consideration of some specific features of magnetic
state of this compound. To reveal and compare paramagnetic effects
of different rare-earth components, the properties of the samples
RuSr$_2$(Eu$_{1.5}$Ce$_{0.5}$)Cu$_{2}$O$_{10-\delta}$ (Ru1222-Eu)
are considered briefly as well.
\section{Results and discussion}
\subsection{Magnetic characterization of the samples}
The samples of Ru1222-Gd and Ru1222-Eu were prepared by a
solid-state reaction method \cite{felner1}. Some of them were set
aside (as-prepared samples), while others were annealed  in pure
oxygen at different pressures. The samples of Ru1222-Gd were
annealed for 12 hours in 30, 62, 78 atm of pure oxygen at
845$^{\circ}$C; whereas, those of Ru1222-Eu were annealed for 24
hours at 800$^{\circ}$C in pure oxygen at pressure of 50 and 100
atm.  The samples have been polycrystalline with a grain size of a
few $\mu$m. They were characterized by resistivity, thermoelectric
power, magnetization and specific-heat measurements, which were in
part reported in Refs.~\cite{don,boris}. It was found there that
superconductivity is affected by granularity and intergrain
Josephson coupling.
\par
In this subsection we present some general magnetic properties of
the samples studied with an emphasis on paramagnetic effects of
rare-earth ions.  The measurements were made with Quantum Design
devices (PPMS and SQUID magnetometer). Temperature behavior of
magnetization, $M(T)$, for the cases of essentially low and
appreciably high magnetic field (Figs. 1 and 2) reveals important
features of complicated magnetic state of these compounds. It is
clearly seen that for both, Ru1222-Gd and Ru1222-Eu, a magnetic
transition takes place when temperature is lowered below
$T_{WF}\approx 90$ K (Fig. 1). This is believed to be the
transition to a weak-ferromagnetic state determined by Ru ions
\cite{felner1,lorenz,awana}. Large difference between the FC and
ZFC curves is likely determined either by high magnetic anisotropy
or spin-glass effects. Magnetic order induced in ruthenocuprates
by RuO$_2$ planes is, however, still unclear
\cite{felner1,lorenz,awana,klamut,knee,lynn,hata,mclau} and will
not be discussed in detail here. We shall dwell briefly only on a
contribution of paramagnetic magnetic moments of rare-earth
components to magnetization of ruthenocuprates.
\par
It is known long ago \cite{vleck,morr} that paramagnetic
properties of trivalent rare-earth ions in chemical compounds are
almost identical to those of quasi-free non-interacting ions. In
both cases paramagnetism is determined by low-lying states of 4$f$
electrons. Effective moment of a rare-earth ion is determined by
quantum numbers $L,S,J$ and according to Hund's rule is
$\mu_{eff}=g[J(J+1)]^{1/2}$, where $g$ is Land\'{e} factor. For
Gd$^{3+}$ ion (ground state ${^8}S_{7/2}$ with $L=0$, $S=J=7/2$)
$\mu_{eff}$ is therefore expected to be 7.94 $\mu_{B}$, in
agreement with experiment \cite{vleck}. For Eu$^{3+}$ ion (ground
state ${^7}F_{0}$ with $L=3$, $S=3$, $J=0$) a significant
deviation from Hund's rule (which predicts the effective moment to
be zero) is found in experiment \cite{vleck}. In particular at
room temperature $\mu_{eff} > 3$ $\mu_{B}$ is observed. The reason
is that for Eu$^{3+}$ ions at high enough temperature the
separation of their ground 4$f$ state (with $J=0$) from higher
level is comparable with $kT$, so that an additional contribution
to susceptibility appears \cite{vleck}. For fairly low
temperature, however, the effective paramagnetic moment for
Eu$^{3+}$ ions is expected to be zero \cite{vleck}.
\par
It can be expected from the aforesaid that Gd$^{3+}$ ions should
give a considerable contribution to the total magnetization of
ruthenocuprates, especially at low temperature; whereas, a
significant contribution of paramagnetic moments of Eu$^{3+}$ is
unlikely. Temperature dependences  of magnetization (Figs. 1 and
2) correspond to the expected behavior. On the whole, specific
magnetization is much higher in the Gd sample as compared with
that of the Eu sample. In particular, $M\approx 10.5$ $\mu_B$/f.u.
at $T=2$~K and $H=7$~T for Ru1222-Gd (Fig. 2). At the same time,
the magnetization of the Ru1222-Eu sample is about 0.9
$\mu_B$/f.u. at the same conditions (Fig. 2).
\par
For both, low and high magnetic fields, $M(T)$ increases as
$T\rightarrow 0$ for the Ru1222-Gd sample, displaying paramagnetic
behavior of Gd ions. In contrast to this, $M(T)$ saturates at low
temperature for the Ru1222-Eu sample. It is evident for the latter
case that the contribution of Eu ions to the total magnetization
is negligible in the low temperature range where the magnitute of
$M$ is determined solely by the magnetic contribution of the Ru
subsystem.
\par
Superconductivity also shows itself to be somewhat different in
the $M(T)$ curves for the Gd and Eu samples. In both samples the
diamagnetic response below the superconducting transition can be
seen in the ZFC curves (Fig. 1), but an appropriate feature in the
FC curve is evident only for Ru1222-Gd sample. With decreasing
temperature when $T$ approaching zero, the diamagnetic response of
the Ru1222-Eu saturates; whereas, that of Ru1222-Gd decreases
(Fig. 1).
\par
Specific-heat measurements have been performed for all of the
Ru1222-Gd samples (as prepared and annealed at different oxygen
pressure). All of these samples have two pronounced features (Fig.
3) in the low-temperature part of $C(T)$ curves: (1) the jump at
the superconducting transition, and (2) the upturn below 20 K
(Schottky-type anomaly). It was found \cite{don,boris} that,
although resistive superconducting transition depends strongly on
intergrain connection determined by oxygen annealing, the position
of the jump in $C(T)$ at the superconducting transition is the
same for all samples studied and reflects in this way the bulk
properties of the compound.
\par
The Eu samples displayed smooth $C(T)$ dependences, which were
identical for all Eu samples studied. The curves were of the Debye
type without any low temperature magnetic anomaly or jump at the
superconducting transition. The former is ascribed to the
non-magnetic nature of Eu ions at low temperature; whereas, the
latter is evidently determined by stronger intergrain disorder in
the Eu samples as compared with the Gd samples \cite{marina}. The
resistive superconducting transitions in the Eu samples are much
broader and normal-state resistivity is approximately ten times
higher than those in the Gd samples \cite{marina}.  It is known
\cite{filler} that a sufficiently strong decoupling between grains
causes smearing and disappearance of the superconducting feature
(jump) in $C(T)$ curves. It should be noted that no feature in the
temperature dependence of the heat capacity, $C(T)$, associated
with magnetic transition at $T\approx 90$~K in the Ru magnetic
subsystem is found in this study. This can be attributed to the
absence of long-range magnetic order in this subsystem at the
transition point due to magnetic inhomogeneities. It is possible
as well that this feature is just too weak to be seen on the
background lattice contribution to specific heat at this rather
high temperature.
\par
The absence of magnetic and superconducting anomalies in $C(T)$
curves for Ru1222-Eu makes it possible to obtain a part of $C(T)$
without lattice contribution \cite{don} by subtraction of the
$C(T)$ curves for Eu from that of the Gd samples, as shown in the
inset of Fig. 3. This shows more clearly the $\lambda$-like
feature at the superconducting transition and the Schottky-type
anomaly below 20 K in Gd sample. The low-temperature Schottky-type
anomaly can be attributed to splitting of the ground term
$^{8}S_{7/2}$ of paramagnetic Gd$^{3+}$ ions by internal and
external magnetic fields, as discussed in more detail in Ref.
\cite{don}.

\subsection{Crossing-point effect}
\par
It is found in this study that curves $C(T)$ for Ru1222-Gd samples
taken at different values of applied magnetic field cross at the
same temperature (the crossing temperature) $T_{\ast} \approx
2.7$~K and specific heat value $C_{\ast}=7.7$~mJ/gK (Fig. 4). This
takes place for each of the Ru1222-Gd samples (as prepared and
annealed at different oxygen pressure). Contrastingly, the $C(T)$
curves of Ru1222-Eu with nonmagnetic Eu ions were found to not
depend on the magnetic field (up to 8 T), as can be expected from
discussion above.
\par
Above the crossing point some clear kink in the $C(T)$ curves
occurs in the temperature range of the Schottky-type anomaly (Fig.
4). This kink is positioned at $T_{k}\approx 5.4$~K for $H=0$, but
with increasing field it shifts to lower temperature and seems to
be smeared for a sufficiently large field (Fig. 4).  In the
low-field $M(T)$ curves (upper panel of Fig. 1), nothing uncommon
can be seen in this temperature range, but in the derivative
$dM/dT$ (Fig.~5), in addition to the very strong features at the
superconducting and magnetic transitions at $T_c$ and $T_{WF}$, a
weak but quite clear peculiarity is seen at $T\simeq 5$~K, that
perhaps pertains to $T_k$.

\par
The appearance and behavior of $T_{k}$ with increasing field is
suggestive of a transition to an antiferromagnetic state for the
Gd$^{3+}$ ion subsystem at low temperature. This type of
transition is ubiquitous in high-$T_c$ cuprates with rare-earth
components \cite{lynn2}. For example, it is found in Gd cuprates
that antiferromagnetic ordering of Gd$^{3+}$ ions takes place at
$T_N$ of 2.3--2.4 K for double CuO$_2$ layer compounds; whereas,
for single-layer ones a higher $T_N$ (up to 6.6 K) is revealed
\cite{ho,lai}. In the related ruthenocuprate RuSr$_2$GdCu$_2$O$_8$
(the Ru1212-type phase) $T_N=2.5$~K is found \cite{lynn3}.
\par
Perhaps the decrease in the superconducting diamagnetic response
with decreasing temperature in Ru1222-Gd (Fig.~1) might be
connected with some kind of magnetic ordering in the Gd$^{3+}$
magnetic subsystem. Unfortunately, the nature of magnetic order
induced in the Ru1222 ruthenocuprates by RuO$_2$ planes is still
not clear \cite{felner1,lorenz,awana,klamut,knee,lynn,hata,mclau}.
This also hampers the determination of the exact nature of the
low-temperature magnetic ordering in the Gd ion subsystem. Another
difficulty is that, up to date only samples of Ru1222 prepared by
solid-state reaction method, have been studied. These samples
usually contain different impurity phases
\cite{knee,lynn,hata,petrykin,asthana}; thus it cannot be ruled
out that the distinct but rather weak feature in $C(T)$ at $T=T_k$
may actually be associated with some magnetic impurity phase. For
example, in the ruthenocuprate
RuSr$_2$(Gd$_{1.3}$Ce$_{0.7}$)Cu$_{2}$O$_{10-\delta}$ (which is
close in composition to that studied in this work) an impurity
phase (5\%) of Sr$_2$GdRuO$_6$ was found \cite{knee} which showed
antiferromagnetic ordering of Gd$^{3+}$ ions near 3 K.
\par
Now let us return again to the crossing point subject. We have
found that in addition to the crossing point at $T_{\ast} \approx
2.7$~K in $C(T)$ curves (taken at different $H$) crossing takes
place also in $C(H)$ curves taken at different temperatures. In
this case the curves cross at $H_{\ast}\approx 3.7$~T (Fig. 6). In
both cases crossing takes place at the same value
$C_{\ast}=7.7$~mJ/gK. Figure 6(b) clearly suggests that $C$ does
not depend on $H$ at $T= T_{\ast} \approx 2.7$~K (dashed line). On
the other hand it is temperature independent at $H=H_{\ast}\approx
3.7$~T (dashed line in Fig. 6(a)). In either case a constant value
of $C$ is $C_{\ast}=7.7$~mJ/gK is observed.
\par
The crossing point effect is considered
\cite{voll,mishra,macedo,eck} as some type of universality for
strongly correlated electron systems, but no unified mechanism for
this phenomenon is proposed. Only some general reasons and
prerequisites for its occurrence have been formulated. It is
believed, for example \cite{voll,mishra,macedo,eck}, that the
crossing ({\it isosbestic}) point occurs in systems which are
close to some quantum  or second-order phase transition, or in
systems with some magnetic instability, so that properties of such
a system are rather sensitive to thermodynamic  variables (like
temperature, pressure, magnetic field).
\par
It is asserted \cite{eck}, among other suggestions, that the
crossing point should become apparent in a system which is a
superposition of two (or more) components, like that in the known
Gorter-Casimir two-fluid model of superconductivity. The total
density of these components, depending, for example, on $T$ and
$H$, is constant,
\begin{equation}
n = n_{1}(T,H) + n_{2}(T,H)=\textrm{const}.
\end{equation}
Following the general concept of such a ``two-fluid'' model
\cite{eck}, some function $f(T,H)$, describing the properties of
this system, can be written as
\begin{equation}
f(T,H) = n_{1}(T,H)f_{1}(H) + n_{2}(T,H)f_{2}(H).
\end{equation}
In this case the crossing point of curves for different
temperatures $T$ should occur at a single point $H_{\ast}$ if
$f_{1}(H_{\ast})= f_{2}(H_{\ast})$. This ``two-fluid'' approach is
perhaps relevant for the crossing point in $C(T,H)$ curves below
the superconducting transition temperature found in the cuprate
Tl$_2$Ba$_2$CuO$_{6+\delta}$\cite{rad}, where the crossing takes
place at $T\approx 0.5$~$T_c$.
\par
In considering of a crossing effect in $C(T,H)$ curves what the
motive force for strong magnetic field dependence of specific heat
is should be first of all taken into account. In the case of Gd
ruthenocuprates considered in this study, the motive force is
connected not with superconductivity, but with splitting of the
ground term $^{8}S_{7/2}$ of paramagnetic Gd$^{3+}$ ions by
internal and external magnetic fields \cite{don}. According to
Kramers' theorem \cite{morr}, the degenerate ground term can be
split into four doublets in tetragonal symmetry.  In particular,
internal molecular fields can arise in the ruthenocuprate from
both the Gd and Ru sublattices and can coexist with
superconductivity. Even though a direct Gd-Gd exchange interaction
is unlikely, these ions can be magnetically polarized by the
4$d$-4$f$ interaction. Generally, the Schottky term in the
specific heat for compounds with Gd$^{3+}$ ions should be
attributed to splitting of all four doublets, although actually
only some of them make the dominant contribution to the effect.
\par
In the simplest case a Schottky term in the specific heat is
determined by properties of a two-level system \cite{kubo}.
Paramagnetic ions in a solid have magnetic dipole moments ($\mu$).
To a first approximation, these do not interact with each other
but can respond to an applied external magnetic field. In a
magnetic field each dipole can exist in one of two states aligned
with the field (spin up) or antialigned (spin down). Spin up
($\uparrow$) and spin down ($\downarrow$) dipoles have an energy
$-\mu H$ and $+\mu H$, respectively. The population of these
discrete energy levels depends on temperature and applied field.
This gives a contribution to the specific heat in a solid known as
Schottky anomaly \cite{kubo}, which is usually seen only at low
temperature, where other contributions are sufficiently small.
\par
It should be mentioned that the total number, $N$, of magnetic
dipoles in the two-level system can be presented as a sum of two
temperature and magnetic-field dependent components $N =
n_{\uparrow}(T,H) + n_{\downarrow}(T,H)=\textrm{const}$, where
$n_{\uparrow}(T,H)$ and $n_{\downarrow}(T,H)$ are numbers of the
spin-up and spin-down dipoles. This relation is similar to
Eq.~(1), so that the ``two-fluid'' approach \cite{eck} is
apparently applicable to a degree to the two-level system as well.
\par
The Schottky-type anomaly in the $C(T,H)$ curves  is by itself
only a background for the crossing effect found in this study,
like that previously seen in NdMnO$_3$ \cite{cheng}. For a deeper
insight into this phenomenon the thermodynamic approach
\cite{voll} can be helpful. It can be suggested rather safely that
within the low temperature range, where the crossing phenomenon
takes place, the magnetic contribution to specific heat is
dominant. The expression for the specific heat at constant $H$ is
\cite{morr} $C_H = T(\partial S/\partial T)_{H}$. Any crossing of
specific heat curves $C_H(T,H)$ means that \cite{voll}
\begin{equation}
\frac{\partial C_{H}(T,H)}{\partial H}\bigg |_{T_{\ast}(H)} =
T_{\ast}(H)\frac{\partial^{2}M(T,H)}{\partial T^2}\bigg
|_{T_{\ast}(H)} =0,
\end{equation}
where the magnetization $M$ is the conjugate thermodynamic
variable for the field $H$. Only if $T_{\ast}$ is independent of
$H$, will all $C_{H}(T,H)$ curves intersect in one point
demonstrating a true crossing effect like that shown in Fig.~6.
\par
It follows from Eq.~(3) that the crossing occurs at the
temperature $T_{\ast}$ where $\partial^{2}M(T,H)/\partial T^2 =0$,
that is $M(T,H)$ must have some type of turning point. Figure~7
shows that $dM/dT$ at $H=7$~T tends to some constant value with
decreasing temperature, while $d^{2}M/dT^{2}$ tends to zero in
this temperature range. We have found that $d^{2}M/dT^{2}$ becomes
approximately zero for $T$ below 3~K. More precise determination
of the point where $d^{2}M/dT^{2}=0$ is difficult due to the
statistical uncertainty of the second derivative. In any case,
however, results of this study support substantially the
theoretical prediction of Ref.~\cite{voll}.
\par
It is seen in Fig. 6 that at $H=H_{\ast}$ specific heat is
temperature independent, that is equal to the constant value
$C_{\ast}=7.7$~mJ/gK,
\begin{equation}
C_{H_{\ast}}(T) = T(\partial S/\partial T)_{H_{\ast}} = C_{\ast}.
\end{equation}
From this it follows that $S_{H_{\ast}} = C_{\ast}\ln T +A$, where
$A$ is a constant. It is also evident that at $H \neq H_{\ast}$
except for the logarithmic term, some polynomial function of $T$
should be added for the approximation of $S_{H}(T)$.
\par
In summary, rather thorough models of {\it isosbestic points} have
been developed for conjugate variables $P$ and $-V$. Some models
were developed for strongly correlated electrons in the frame of
the Hubbard model \cite{voll, mishra,macedo,eck}. We hope that
results of this study will promote development of an adequate
model for crossing point in $C(T,H)$ curves for magnetic systems
undergoing a transition from classical to quantum behavior in
$C(T,H)$ with decreasing temperature.
\par
Work at Texas A\&M University was supported by Grant A-0514 from
the Robert A Welch Foundation.



\newpage

\centerline{\bf{Figures}} \vspace{12pt}

Fig.~1. Temperature behavior of specific magnetization (field
cooled (FC) and zero-field cooled (ZFC) curves) at field $H
=0.5$~mT for samples Ru1222-Gd
(RuSr$_2$(Gd$_{1.5}$Ce$_{0.5}$)Cu$_{2}$O$_{10-\delta}$, annealed
12 hours at 845$^{\circ}$C in pure oxygen at pressure 78 atm) and
Ru1222-Eu (RuSr$_2$(Eu$_{1.5}$Ce$_{0.5}$)Cu$_{2}$O$_{10-\delta}$,
as-prepared state). Temperature of the intragrain superconducting
transition $T_{c}\approx 34$ K is indicated by an arrow on $M(T)$
curve for Ru1222-Gd. The temperature $T_{WF}$, for the presumed
transition to a weak-ferromagnetic state in the Ru magnetic
subsystem is marked by arrows for both samples. Other features of
the $M(T)$ curves are discussed in the main text. \vspace{15pt}

Fig.~2. Temperature behavior of magnetization  (in $\mu_{B}$ per
formula unit) at field $H = 7$~T for samples Ru1222-Gd
(RuSr$_2$(Gd$_{1.5}$Ce$_{0.5}$)Cu$_{2}$O$_{10-\delta}$, annealed
12 hours at 845$^{\circ}$C in pure oxygen at pressure 78 atm) and
Ru1222-Eu (RuSr$_2$(Eu$_{1.5}$Ce$_{0.5}$)Cu$_{2}$O$_{10-\delta}$,
annealed 24 hours at 800$^{\circ}$C in pure oxygen at pressure 100
atm).\vspace{15pt}

Fig.~3. (Color online) Low-temperature behavior of total specific
heat at zero magnetic field for samples Ru1222-Gd
(RuSr$_2$(Gd$_{1.5}$Ce$_{0.5}$)Cu$_{2}$O$_{10-\delta}$, annealed
12 hours at 845$^{\circ}$C in pure oxygen at pressure 78 atm) and
Ru1222-Eu (RuSr$_2$(Eu$_{1.5}$Ce$_{0.5}$)Cu$_{2}$O$_{10-\delta}$,
annealed 24 hours at 800$^{\circ}$C in pure oxygen at pressure 100
atm). Difference between the $C(T)$ curves of Ru1222-Gd and
Ru1222-Eu (inset) shows more clearly the $\lambda$-like feature at
the superconducting transition and a Schottky-type anomaly below
20 K.  \vspace{15pt}

Fig.~4. (Color online) Low-temperature dependences of total
specific heat, $C$, taken at different values of applied magnetic
fields for two samples of
RuSr$_2$(Gd$_{1.5}$Ce$_{0.5}$)Cu$_{2}$O$_{10-\delta}$: (a)
annealed 12 hours at 845$^{\circ}$C in pure oxygen at pressure 78
atm, and (b) as-prepared state. In both cases the $C(T)$ curves
cross at the same temperature $T_{\ast} \approx 2.7$~K (at the
same specific heat value $C_{\ast}=7.7$~mJ/gK), revealing the
crossing point phenomenon. Arrows indicate temperature $T_k$ of a
kink in $C(T)$ curves, which is about 5.4~K for $H=0$ and moves to
lower temperature with increasing field.\vspace{15pt}

Fig.~5. Derivative $dM/dT$ (of the FC $M(T)$ curve at $H=0.5$~mT,
shown in Fig.~1)  for the sample of
RuSr$_2$(Gd$_{1.5}$Ce$_{0.5}$)Cu$_{2}$O$_{10-\delta}$, annealed 12
hours at 845$^{\circ}$C in pure oxygen at pressure 78 atm. The
arrows indicate temperatures $T_{WF}$, $T_c$ and $T_k$ of phase
transitions discussed in the text.\vspace{15pt}

Fig.~6. (Color online) Temperature and magnetic field dependences
of specific heat, $C$, taken, respectively, at different values of
applied magnetic field (a) or temperature (b). All data shown are
for the sample of
RuSr$_2$(Gd$_{1.5}$Ce$_{0.5}$)Cu$_{2}$O$_{10-\delta}$, annealed 12
hours at 845$^{\circ}$C in pure oxygen at a pressure 78 atm,
except the data for $H=8$~T, which is taken for the as-prepared
sample. The $C(T)$ or $C(H)$ curves cross at the temperature
$T_{\ast} \approx 2.7$~K (a) or magnetic field $H_{\ast}=3.7$~T
(b) for the same value $C_{\ast}=7.7$~mJ/gK, demonstrating the
crossing point phenomenon.\vspace{15pt}

Fig.~7. (Color online) Derivatives $dM/dT$ and $d^{2}M/dT^{2}$ of
the $M(T)$ curve at $H=7$~T (shown in Fig.~2) for sample of
RuSr$_2$(Gd$_{1.5}$Ce$_{0.5}$)Cu$_{2}$O$_{10-\delta}$, annealed 12
hours at 845$^{\circ}$C in pure oxygen at pressure 78 atm.

\newpage
\begin{figure}
\centering\includegraphics[width=0.8\linewidth]{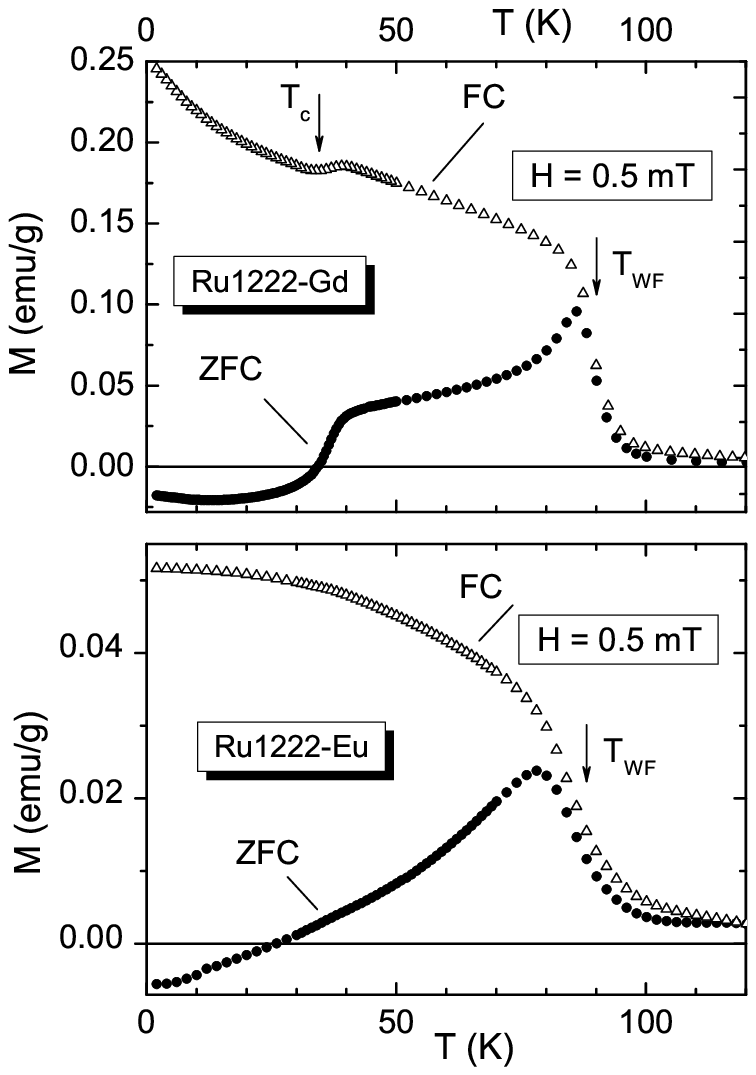}
\centerline{Figure 1 to paper Belevtsev et al.}
\end{figure}

\begin{figure}
\centering\includegraphics[width=0.8\linewidth]{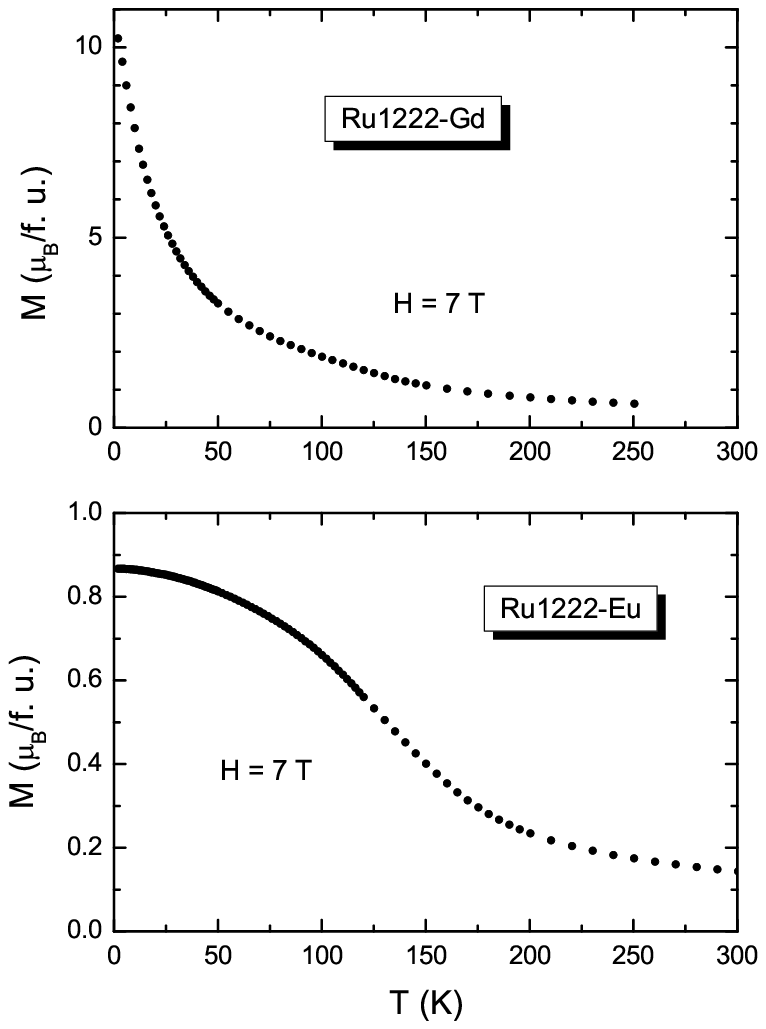}
\centerline{Figure 2 to paper Belevtsev et al.}
\end{figure}

\begin{figure}
\centering\includegraphics[width=0.8\linewidth]{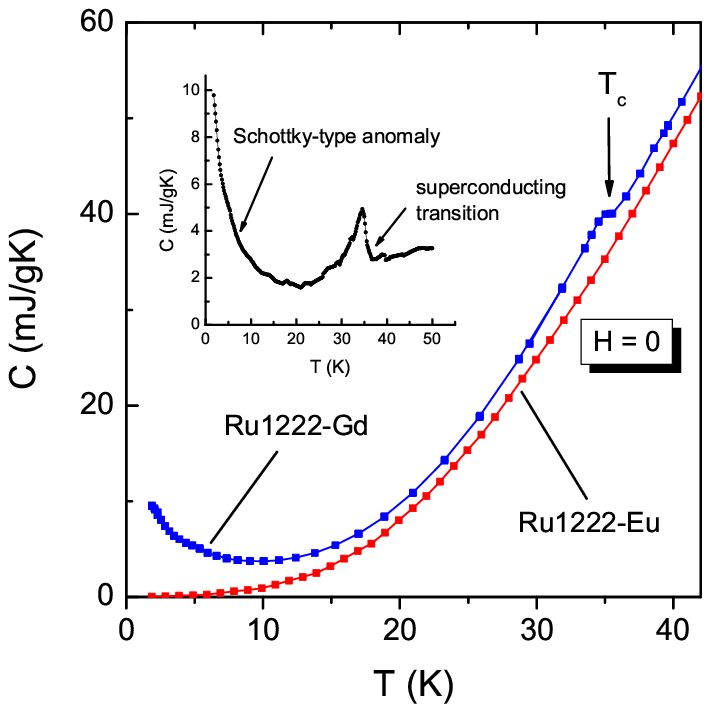}
\centerline{Figure 3 to paper Belevtsev et al.}
\end{figure}

\begin{figure}
\centering\includegraphics[width=0.8\linewidth]{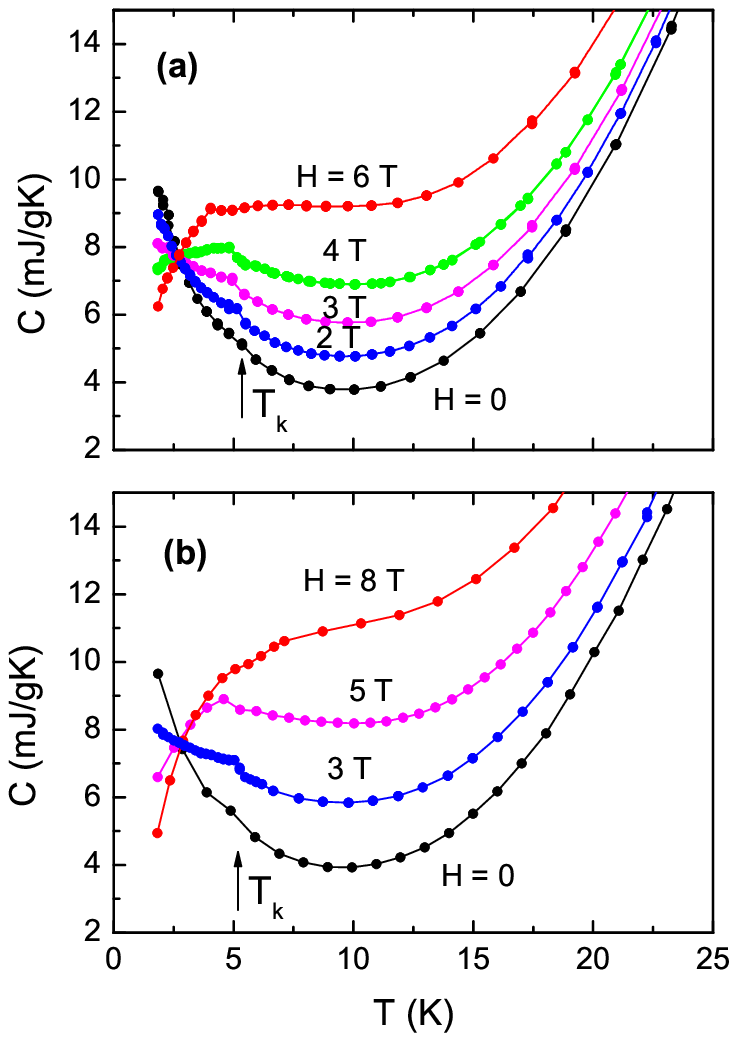}
\centerline{Figure 4 to paper Belevtsev et al.}
\end{figure}

\begin{figure}
\centering\includegraphics[width=0.8\linewidth]{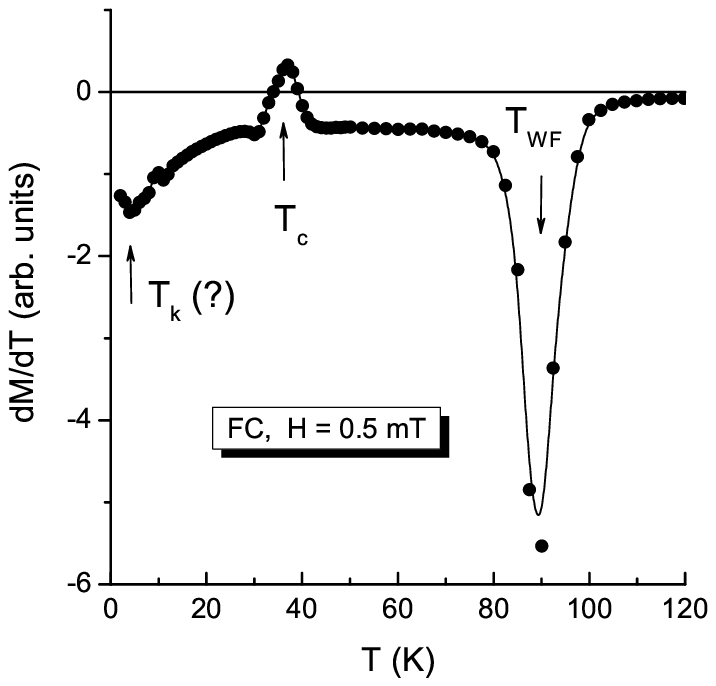}
\centerline{Figure 5 to paper Belevtsev et al.}
\end{figure}

\begin{figure}
\centering\includegraphics[width=0.8\linewidth]{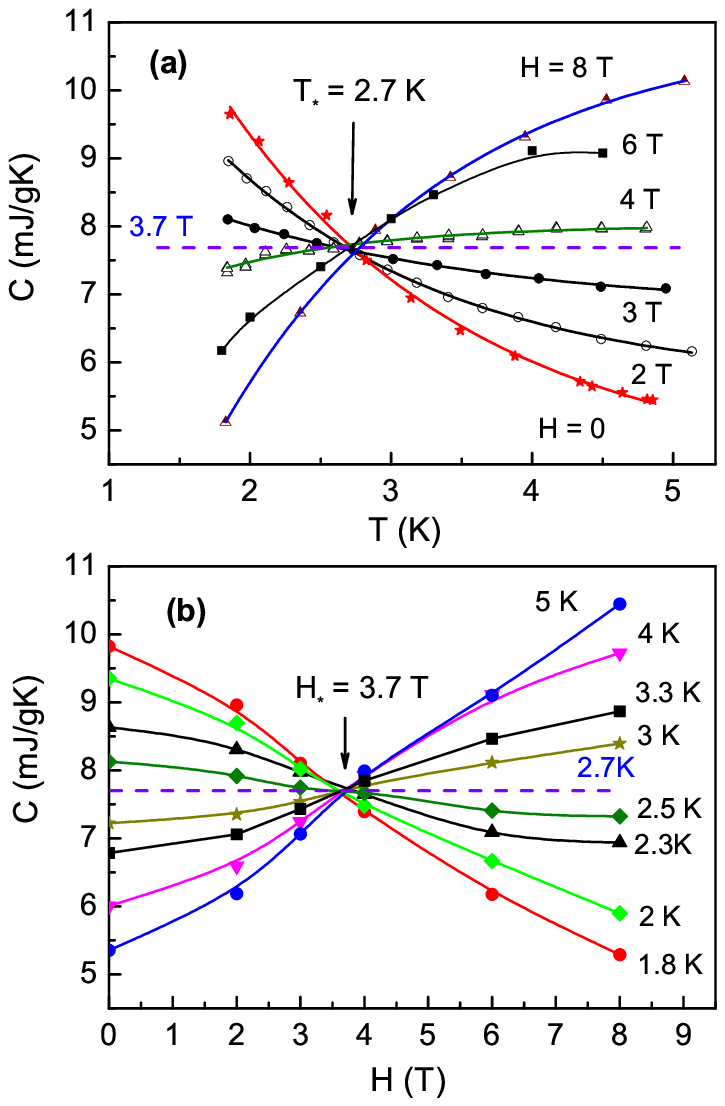}
\centerline{Figure 6 to paper Belevtsev et al.}
\end{figure}

\begin{figure}
\centering\includegraphics[width=0.8\linewidth]{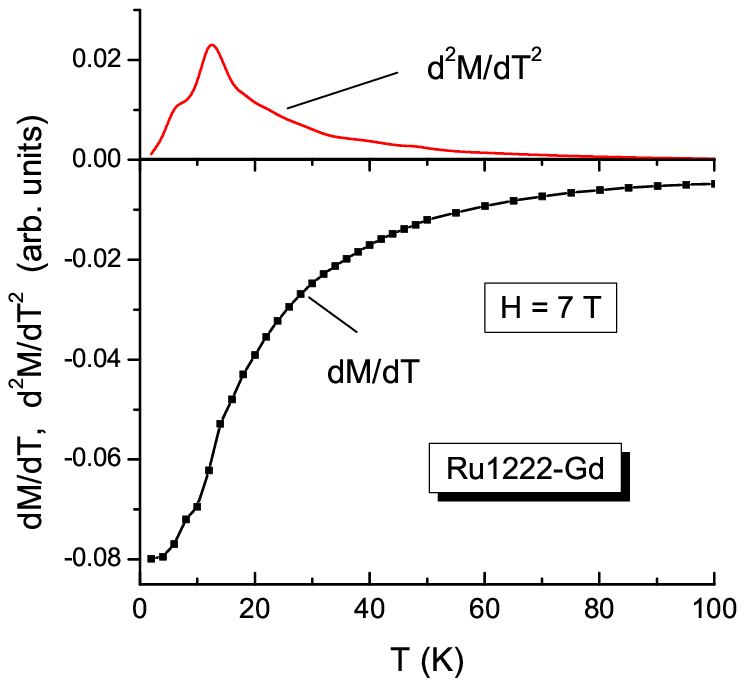}
\centerline{Figure 7 to paper Belevtsev et al.}
\end{figure}


\section*{References}

\begin{thebibliography}{00}

\bibitem{voll}Vollhardt D 1997 {\it Phys. Rev. Lett.} {\bf 78} 1307

\bibitem{mishra}Mishra S G and Sreeram P A 2000 {\it Eur. Phys. J. B}
{\bf 14} 287

\bibitem{macedo}Macedo  C A and  de Souza A M C 2002 {\it Phys. Rev.
B} {\bf 65} 153109

\bibitem{eck} Eckstein Martin, Kollar Marcus and Vollhardt
Dieter 2007 {\it J. Low Temp. Phys.} {\bf 147} 279

\bibitem{schlag}Schlager H G, Schr\"{o}der A, Welsch M and
L\"{o}hneysen H von 1993 {\it J. Low Temp. Phys.} {\bf 90} 181

\bibitem{fisch}Fischer J, Schr\"{o}der A, L\"{o}hneysen H von,
Bauhofer W and Steigenberger U 1989 {\it Phys. Rev. B} {\bf 39}
11775

\bibitem{ho}Ho J C, Chen Y Y, Yao Y D, Huang W S, Sheen S R,
Huang J C and Wu M K 1997 {\it Physica C} {\bf 282-287} 1403

\bibitem{cheng}Cheng J G, Sui Y, Qian Z N, Liu Z G, Miao J P,
Huang X Q, Lu Z, Li Y, Li Y, Wang X J and Su W H 2005 {\it Solid
State Commun.} {\bf 134} 381

\bibitem{felner1} Felner I 2003 {\it Studies of High Temperature
Superconductors} vol~46 ed A V Narlikar (New York: Nova Science)
pp 41-75

\bibitem{lorenz}Lorenz B, Xue Y Y and Chu C W 2003 {\it Studies of High Temperature
Superconductors} vol~46 ed A V Narlikar (New York: Nova Science)
pp 1-39

\bibitem{awana}Awana V. P. S. 2005 {\it Frontiers in magnetic Materials}
ed A V Narlikar (Berlin: Springer) pp 531-571 (also in {\it
Preprint} cond-mat/0407799)

\bibitem{klamut}Klamut P W 2008 {\it Supercond. Sci. Technol.} {\bf 21}
093001

\bibitem{don}Naugle D G, Rathnayaka K D D, Krasovitsky V B,
Belevtsev B I, Anatska M P, Agnolet G  and Felner I 2006 {\it J.
Appl. Phys.} {\bf 99} 08M501

\bibitem{boris}Belevtsev B I, Beliayev E Yu, Naugle D G,
Rathnayaka K D D, Anatska M P and Felner I 2007 {\it J. Phys.:
Condens. Matter} {\bf 19} 036222

\bibitem{knee}Knee C S, Rainford B D and Weller M T 2000 {\it J. Mater.
Chem.} {\bf 10} 2445

\bibitem{lynn}Lynn J W, Chen Y, Huang Q, Goh S K and  Williams G V
M 2007 {\it Phys. Rev. B} {\bf 76} 014519

\bibitem{hata} Hata Y, Uragami Y, Yasuoka H 2008 {\it Physica C} {\bf 468} 2392

\bibitem{mclau}Mclaughlin A C, Felner I and Awana V P S 2008
{\it Phys. Rev. B} {\bf 78} 094501

\bibitem{vleck}Van Vleck J H 1932 {\it The Theory of Electric and Magnetic
Susceptibilities} (London: Oxford University Press)

\bibitem{morr} Morrish Allan H 1965 {\it The Physical Principles of Magnetism}
 (New York: John Wiley \& Sons)

\bibitem{marina} Anatska M P 2006 {\it The transport coefficients in
(R$_{1.5}$Ce$_{0.5}$)RuSr$_2$Cu$_2$O$_{10-\delta}$ (R=Gd,Eu)
rutheno-cuprates} M. S. Thesis (College Station, Texas: Texas A\&M
University) http://txspace.tamu.edu/handle/1969.1/5019?show=full

\bibitem{filler} Filler R L, Lindenfeld P, Worthington T and Deutscher G
1980 {\it Phys. Rev. B} {\bf 21} 5031

\bibitem{lynn2}Lynn J W and Skanthakumar S 2001 {\it Handbook on
the Physics and Chemistry of Rare Earths} ed K A Gschneider Jr, L
Eyring and M B Maple (Amsterdam: North Holland) Vol 31, Chap 199,
pp 315-350

\bibitem{lai} Lai C C, Shieh J H, Chiou B S, Ho J C and Ku H C
1994 {\it Phys. Rev. B} {\bf 49} 1499

\bibitem{lynn3}Lynn J W, Keimer B, Ulrich C, Bernhard C
and Tallon J L 2000 {\it Phys. Rev. B} {\bf 61} R14964

\bibitem{petrykin}Petrykin V V, Osada M, Kakihana M, Tanaka Y,
Yasuoka H, Ueki Y, Abe M, 2003 {\it Chem. Mater.} {\bf 15} 4417

\bibitem{asthana}Asthana Anjana and Matsui Yoshio 2008 {\it Physica C}
 {\bf 468} 458

\bibitem{rad}Radcliffe J W, Loram J W, Wade J M, Wltschek G
and Tallon J L 1996 {\it J. Low Temp. Phys.} {\bf 105} 903

\bibitem{kubo}Kubo Ryogo 1965 {\it Statistical Mechanics} (Amsterdam:
North-Holland Publishing Company)


\end{thebibliography}
\end{document}